\documentstyle[12pt,sw20lart]{article}

\renewcommand{\thesection}{\arabic{section}} 
\input tcilatex
\QQQ{Language}{
American English
}

\begin{document}

\title{On mass-dependent subtraction and removal of the ambiguity in QED
renormalization }
\author{Jun-Chen Su, Xue-Xi Yi and Ying-Hui Cao \\
Center for Theoretical Physics, Department of Physics,\\
Jilin University, Changchun 130023, \\
People's Republic of China}
\date{}
\maketitle

\begin{abstract}
~~ The QED renormalization is restudied by using a mass-dependent
subtraction which is performed at a time-like renormalization point. The
subtraction exactly respects necessary physical and mathematical
requirements such as the gauge symmetry, the Lorentz- invariance and the
mathematical convergence. Therefore, the renormalized results derived in the
subtraction scheme are faithful and have no ambiguity. Especially, it is
proved that the solution of the renormalization group equation satisfied by
a renormalized wave function, propagator or vertex can be fixed by applying
the renormalization boundary condition and, thus, an exact S-matrix element
can be expressed in the form as written in the tree diagram approximation
provided that the coupling constant and the fermion mass are replaced by
their effective ones. In the one-loop approximation, the effective coupling
constant and the effective fermion mass obtained by solving their
renormalization group equations are given in rigorous and explicit
expressions which are suitable in the whole range of distance and exhibit
physically reasonable asymptotic behaviors.
\end{abstract}

\vskip 1truecm

\setcounter{section}{1}

\section*{1.Introduction}

~~ In the renormalization of quantum field theories, to extract finite
physical results from higher order perturbative calculations, a certain
subtraction scheme is necessary to be used so as to remove the divergences
occurring in the calculations. There are various subtraction schemes in the
literature, such as the minimal subtraction (MS)$^1$, the modified minimal
subtraction ($\overline{MS}$) $^2$, the momentum space subtraction (MOM)$^3$%
, the on-mass-shell subtraction (OS) $^{4,5}$, the off-mass shell
subtraction (OMS)$^6$ and etc.. However, there exists a serious ambiguity
problem$^{3,6.7}$ that different subtraction schemes in general give
different physical predictions, conflicting the fact that the physical
observables are independent of the subtraction schemes. Ordinarily, it is
argued that the ambiguity appears only in finite order perturbative
calculations, while the exact result given by the whole perturbation series
is scheme- independent. Though only a finite order perturbative calculation
is able to be done in practice, one still expects to get unambiguous results
from such a calculation. To solve the ambiguity problem, several
prescriptions, such as the minimal sensitivity principle$^7$, the effective
charge method$^8$ and some others$^9$ , were proposed in the past. By the
principle of minimal sensitivity, an additional condition has to be
introduced and imposed on the result calculated in a finite order
perturbative approximation so as to obtain an optimum approximant which is
least sensitive to variations in the unphysical parameters. In the effective
charge method, the use of a coupling constant is abandoned. Instead, an
effective charge is associated with each physical quantity and used to
determine the Gell-Mann-Low function in the renormalization group equation
(RGE). In this way, a renormalization-scheme-invariant result can be found
from the RGE. In Ref.(9), the authors developed a new perturbation approach
to renormalizable field theories. This approach is based on the observation
that if the perturbation series of a quantity R is not directly computable
due to that the expansion coefficients are infinite, the unambiguous result
of the quantity can be found by solving the differential equation $Q\frac{%
dR(Q)}{dQ}=F(R(Q))$ where F(R) is well-defined and can be expanded as a
series of R. Particularly, the finite coefficients of the expansion of F(R)
are renormalization-scheme-independent and in one-to-one correspondence with
the ones in the ordinary perturbation series of the R. The prescriptions
mentioned above are somehow different from the conventional perturbation
theory in which the coupling constant is commonly chosen to be the expansion
parameter for a perturbation series.

In this paper, we wish to deal with the ambiguity problem from a different
angle. It will be shown that the ambiguity problem can be directly tackled
in the conventional perturbation theory by the renormalization group method$%
^{10-15}$. The advantage of this method is that the anomalous dimension in a
RGE is well-defined although it is computed from the renormalization
constant which is divergent in its original definition. In comparison of the
renormalization group method with the aforementioned approach proposed in
Ref.(9), we see, both of them are much similar to one another in
methodology. This suggests that the renormalization group method is also
possible to yield the theoretical results which are free from the ambiguity.
The possibility relies on how to choose a good subtraction scheme which
gives rise to such renormalization constants that they lead to unique
anomalous dimensions. The so-called good subtraction scheme means that it
must respect necessary physical and mathematical requirements such as the
gauge symmetry, the Lorentz invariance and the mathematical convergence
principle. The necessity of these requirements is clear. Particularly, in
the renormalization group approach to the renormalization problem, the
renormalization constants obtained from a subtraction scheme not only serve
to subtract the divergences, but also are directly used to derive physical
results. Apparently, to guarantee the calculated results to be able to give
faithful theoretical predictions, the subtraction procedure necessarily
complies with the basic principles established well in physics and
mathematics. Otherwise, the subtraction scheme should be discarded and thus
the scheme-ambiguity will be reduced. Let us explain this viewpoint in some
detail from the following aspects: (1) For a gauge field theory, as one
knows, the gauge-invariance is embodied in the Ward identity. This identity
is a fundamental constraint for the theory. Therefore, a subtraction scheme,
if it is applicable, could not defy this identity. As will be demonstrated
in latter sections, the Ward identity not only establishes exact relations
between renormalization constants, but also determines the functional
structure of renormalization constants; (2) The Lorentz-invariance is partly
reflected in the energy-momentum conservation which holds at every vertex.
Since the renormalization point in the renormalization constants will be
eventually transformed to the momentum in the solutions of RGEs, obviously,
in order to get correct functional relations of the solutions with the
momentum, the energy-momentum conservation could not be violated by the
subtraction of vertices; (3) Why the convergence principle is required in
the renormalization calculation? As we know, the renormalization constants
are divergent in their original appearance. Such divergent quantities are
not well-defined mathematically and hence are not directly calculable$^9$%
because we are not allowed to apply any computational rule to do a
meaningful or unambiguous calculation for this kind of quantities. The
meaningful calculation can only be done for the regularized form of
renormalization constants which are derived from corresponding regularized
Feynman integrals in a subtraction scheme. The necessity of introducing the
regularization procedure in the quantum field theory may clearly be seen
from the mathematical viewpoint as illustrated in Appendix A. Therefore, in
renormalization group calculations, the correct procedure of computing
anomalous dimensions is starting from the regularized form of
renormalization constants. The limit operation taken for the regularization
parameter should be performed after completing the differentiation with
respect to the renormalization point. Since the anomalous dimensions are
convergent, the limit is meaningful and would give definite results.
Obviously, the above procedure of computing the anomalous dimensions agrees
with the convergence principle; (4) In comparison with the other subtraction
schemes, the MOM scheme appears to be more suitable for renormalization
group calculations. This is because this scheme naturally provides not only
a renormalization point which is needed for the renormalization group
calculation, but also a renormalization boundary condition for a
renormalized quantity (a wave function, a vertex or a propagator), which
will be used to fix the solution of the RGE for the renormalized quantify.

Based on the essential points of view stated above, it may be found that a
renormalized quantity can be unambiguously determined by its RGE and thus a
renormalized S-matrix element can be given in an unique form without any
ambiguity. To illustrate this point, we limit ourselves in this paper to
take the QED renormalization as an example to show how the ambiguity can be
eliminated. As one knows, the QED renormalization has been extensively
investigated by employing the OS, $\overline{MS}$ and MOM schemes in the
previous works$^{4-6,10-26}$. But, most of these studies are concentrated on
the large momentum (short distance) behaviors of some quantities for the
sake of simplicity of the calculation. In this paper, we restudy the QED
renormalization with the following features: (1) The renormalization is
performed in a mass-dependent scheme other than in the mass-independent
scheme which was adopted in many previous works$^{20-26}$. The
mass-dependent scheme obviously is more suitable for the case that the mass
of a charged fermion can not be set to be zero; (2) The subtraction is
carried out in such a MOM scheme that the renormalization point is mainly
taken to be an arbitrary time-like momentum other than a space-like momentum
as chosen in the conventional MOM scheme. The time-like MOM scheme actually
is a generalized mass-shell scheme (GMS). The prominent advantage of the GMS
scheme is that in this scheme the scale of renormalization point can
naturally be connected with the scale of momenta and the results obtained
can directly be converted to the corresponding ones given in the OS-scheme ;
(3) The subtraction is implemented by fully respecting the necessary
physical and mathematical principles mentioned before . Therefore, the
results obtained are faithful and free of ambiguity; (4) The effective
coupling constant and the effective fermion mass obtained in the one-loop
approximation are given exact and explicit expressions which were never
found in the literature. These expressions exhibit physically reasonable
infrared and ultraviolet behaviors. We will pay main attention to the
infrared (large distance) behavior because this behavior is more sensitive
to identify whether a subtraction is suitable or not for the QED
renormalization.

~~The rest of this paper is arranged as follows. In Sect.2, we sketch the
RGE and its solution and show how a S-matrix element can be free of
ambiguity. In Sect.3, we briefly discuss the Ward identity and give a
derivation of the subtraction version of the fermion self-energy in the GMS
scheme. In Sect.4, we derive an exact expression of the one-loop effective
coupling constant and discuss its asymptotic property. In Sect.5, the same
thing will be done for the effective fermion mass. The last section serves
to make some comments and discussions. In Appendix A, we show a couple of
mathematical examples to help understanding the regularization procedure
used in the renormalization group calculations.

\setcounter{section}{2}

\section*{2.Solution to RGE and S-matrix element}

\setcounter{equation}{0}

~~ Among different formulations of the RGE (see the review given in Ref.(14))%
$^{10-15}$, we like to employ the approach presented in Ref.(15). But, we
work in a mass-dependent renormalization scheme, therefore, the anomalous
dimension in the RGE depends not only on the coupling constant, but also on
the fermion mass. Suppose $F_R$ is a renormalized quantity. In the
multiplicative renormalization, it is related to the unrenormalized one $F$
in such a way 
\begin{eqnarray}
F=Z_FF_R
\end{eqnarray}
where $Z_F$ is the renormalization constant of $F$. In GMS scheme, the $Z_F$
and $F_R$ are all functions of the renormalization point $\mu =\mu _0e^t$.
Differentiating Eq.(2.1) with respect to the $\mu $ and noticing that the $F$
is independent of $\mu $, we immediately obtain a RGE satisfied by the
function $F_R^{}$ 
\begin{eqnarray}
\mu \frac{dF_R}{d\mu }+\gamma _FF_R=0
\end{eqnarray}
where $\gamma _F$ is the anomalous dimension defined by 
\begin{eqnarray}
\gamma _F=\mu \frac d{d\mu }\ln Z_F
\end{eqnarray}
We first note here that the anomalous dimension can only depends on the
ratio ${\sigma =\frac{m_R}\mu }$, ${\gamma }_F{=\gamma }_F{(g}_R{,\sigma ),}$
because the renormalization constant is dimensionless. Next, we note,
Eq.(2.2) is suitable for a physical parameter (mass or coupling constant), a
propagator, a vertex, a wave function or some other Green function. If the
function ${F_R}$ stands for a renormalized Green function, vertex or wave
function, in general, it not only depends explicitly on the scale $\mu $,
but also on the renormalized coupling constant $g_R$, mass $m_R$ and gauge
parameter $\xi _R$ which are all functions of $\mu $, $F_R=F_R(p,g_R(\mu
),m_R(\mu ),\xi _R(\mu );\mu )$ where $p$ symbolizes all the momenta.
Considering that the function ${F_R}$ is homogeneous in the momentum and
mass, it may be written, under the scaling transformation of momentum $%
p=\lambda p_0$ , as follows 
\begin{eqnarray}
F_R(p;g_R,m_R,\xi _R,\mu )=\lambda ^{D_F}F_R(p_0;g_R,\frac{m_R}\lambda ,\xi
_R;\frac \mu \lambda )
\end{eqnarray}
where $D_F$ is the canonical dimension of $F$. Since the renormalization
point is a momentum taken to subtract the divergence, we may set $\mu =\mu
_0\lambda $ where $\lambda =e^t$ which is taken to be the same as in $%
p=p_0\lambda $. Noticing the above transformation, the solution of the RGE
in Eq.(2.2) can be expressed as$^{15}$ 
\begin{eqnarray}
F_R(p;g_R,m_R,\xi _R,\mu _0) &=&\lambda ^{D_F}e^{\int_1^\lambda \frac{%
d\lambda }\lambda \gamma _F(\lambda )}F_R(p_0;  \nonumber \\
&&g_R(\lambda ),m_R(\lambda )\lambda ^{-1},\xi _R(\lambda );\mu _0)
\end{eqnarray}
where $g_R(\lambda ),m_R(\lambda )$ and $\xi _R(\lambda )$ are the running
coupling constant, the running mass and the running gauge parameter,
respectively. The solution written above shows the behavior of the function $%
F_R$ under the scaling of momenta.

How to determine the function $F_R(p_0;\cdots ,\mu _0)$ on the RHS of
Eq.(2.5) when the $F_R(p_0,...)$ stands for a wave function, a propagator or
a vertex? This question can be unambiguously answered in MOM scheme, but was
not answered clearly in the literature $^{27,28}$. Noticing that the
momentum $p_0$ and the renormalization point $\mu _0$ are fixed, but may be
chosen arbitrarily, we may, certainly, set $p_0^2=\mu _0^2$. With this
choice, by making use of the following boundary condition satisfied by a
propagator, a vertex or a wave function 
\begin{equation}
F_R(p_0;g_R,m_R,\xi _R,\mu )\mid _{P_0^2=\mu ^2}=F_R^{(0)}(p_0;g_R,m_R,\xi
_R)
\end{equation}
where the function $F_R^{(0)}(p;g_R,m_R,\alpha _R)$ is of the form of free
propagator, bare vertex or free wave function and independent of the
renormalization point (see the examples given in the next section) and
considering the homogeneity of the function $F_R$ as mentioned in Eq.(2.4),
we may write 
\begin{eqnarray}
\lambda ^{D_F}F_R &(p_0;g_R(\lambda ),m_R(\lambda )\lambda ^{-1},\xi
_R(\lambda ),\mu _0)\mid &_{p_0^2=\mu _0^2}  \nonumber \\
=F_R^{(0)} &&(p;g_R(\lambda ),m_R(\lambda ),\xi _R(\lambda ))
\end{eqnarray}
where the renormalized coupling constant, mass and vertex in the function $%
F_R^{(0)}(p,...)$ become the running ones. With the expression given in
Eq.(2.7), Eq.(2.5) will finally be written in the form 
\begin{eqnarray}
F_R(p;g_R,m_R,\xi _R)=e^{\int_1^\lambda \frac{d\lambda }\lambda \gamma
_F(\lambda )}F_R^{(0)}(p;g_R(\lambda ),m_R(\lambda ),\xi _R(\lambda ))
\end{eqnarray}
For a gauge field theory, it is easy to check that the anomalous dimension
in Eq.( 2.8) will be cancelled out in S-matrix elements. To show this point
more specifically, let us take the two-electron scattering taking place in
t-channel as an example. Considering that a S-matrix element expressed in
terms of unrenormalized quantities is equal to that represented by the
corresponding renormalized quantities, the scattering amplitude may be
written as 
\begin{equation}
S_{fi}=\overline{u}_R^{\alpha ^{^{\prime }}}(p_1^{^{\prime }})\Gamma _R^\mu
(p_1^{^{\prime }},p_1)u_R^\alpha (p_1)iD_{R\mu \nu }(k)\overline{u}_R^{\beta
^{^{\prime }}}(p_2^{^{\prime }})\Gamma _R^\nu (p_2^{^{\prime
}},p_2)u_R^\beta (p_2)
\end{equation}
where $k=p_1^{\prime }-p_1=p_2^{}-p_2^{\prime }$; $u_R^\alpha (p)$, $\Gamma
_R^\mu (p^{\prime },p)$ and $iD_{R\mu \nu }(k)$ represent the fermion wave
function, the proper vertex and the photon propagator respectively which are
all renormalized. The renormalization constants of the wave function, the
propagator and the vertex will be designated by $\sqrt{Z_2},Z_3$ and $%
Z_\Gamma $ respectively. The constant $Z_\Gamma $ is defined as 
\begin{equation}
Z_\Gamma =Z_2^{-1}Z_3^{-\frac 12}
\end{equation}
because the vertex in Eq.(2.9) contains a coupling constant in it.

According to the formula given in Eq.(2.8), the renormalized fermion wave
function, photon propagator and vertex can be represented in the forms as
shown below. For the fermion wave function, we have 
\begin{equation}
u_R^\alpha (p)=e^{\int_1^\lambda \frac{d\lambda }\lambda \gamma _F(\lambda
)}u_{R\alpha }^{(0)}(p,m_R(\lambda ))
\end{equation}
where 
\begin{equation}
u_{R\alpha }^{(0)}(p,m_R(\lambda ))=\left( \frac{E+m_R(\lambda )}{%
2m_R(\lambda )}\right) ^{\frac 12}\left( \frac{\overrightarrow{\sigma }.%
\overrightarrow{p}}{E+m_R(\lambda )}\right) \varphi _\alpha (\overrightarrow{%
p})
\end{equation}
is the free wave function in which $m_R(\lambda )$ is the running mass and 
\begin{equation}
\gamma _F=\frac 12\mu \frac d{d\mu }\ln Z_2
\end{equation}
is the anomalous dimension of fermion wave function. For the renormalized
photon propagator, we can write 
\[
iD_{R\mu \nu }(k)=e^{\int_1^\lambda \frac{d\lambda }\lambda \gamma
_3(\lambda )}iD_{R\mu \nu }^{(0)}(k) 
\]
where 
\begin{equation}
iD_{R\mu \nu }^{(0)}(k)=-\frac i{k^2+i\varepsilon }[g_{\mu \nu }-(1-\alpha
_R(\lambda ))\frac{k_\mu k_\nu }{k^2}]_{}
\end{equation}
is the free propagator with $\alpha _R(\lambda )$ being the running gauge
parameter in it and 
\begin{equation}
\gamma _3(\lambda )=\mu \frac d{d\mu }\ln Z_3
\end{equation}
is the anomalous dimension of the propagator. For the renormalized vertex,
it reads 
\begin{equation}
\Gamma _R^\mu (p^{\prime },p)=e^{\int_1^\lambda \frac{d\lambda }\lambda
\gamma _\Gamma (\lambda )}\Gamma _R^{(0)\mu }(p^{\prime },p)
\end{equation}
where

\begin{equation}
\Gamma _R^{(0)\mu }(p^{\prime },p)=ie_R(\lambda )\gamma ^\mu
\end{equation}
is the bare vertex containing the running coupling constant (the electric
charge) $e_R(\lambda )$ in it and 
\begin{equation}
\gamma _\Gamma (\lambda )=\mu \frac d{d\mu }\ln Z_\Gamma =-\mu \frac d{d\mu
}\ln Z_2-\frac 12\mu \frac d{d\mu }\ln Z_3
\end{equation}
is the anomalous dimension of the vertex here the relation in Eq.(2.10) has
been used. Upon substituting Eqs.(2.11), (2.14) and (2.17) into Eq.(2.9) and
noticing Eqs.( 2.13), (2.16) and (2.19), we find that the anomalous
dimensions in the S-matrix element are all cancelled out with each other. As
a result, we arrive at 
\begin{equation}
S_{fi}=\overline{u}_{R\alpha ^{^{\prime }}}^{(0)}(p_1^{\prime })\Gamma
_R^{(0)\mu }(p_1^{\prime },p_1)u_{R\alpha }^{(0)}(p_1)iD_{R\mu \nu }^{(0)}(k)%
\overline{u}_{R\beta ^{^{\prime }}}^{(0)}(p_2^{\prime })\Gamma _R^{(0)\nu
}(p_2^{\prime },p_2)u_{R\beta }^{(0)}(p_2)
\end{equation}
This expression clearly shows that the exact S-matrix element of the
two-electron scattering can be represented in the form as given in the
lowest order (tree diagram) approximation except that all the physical
parameters in the matrix elements are replaced by their effective (running)
ones. For other S-matrix elements, the conclusion is the same. This is
because any S-matrix element is unexceptionably expressed in terms of a
number of wave functions, propagators and proper vertices each of which can
be represented in the form as shown in Eq.(2.8) and the anomalous dimensions
in the matrix element , as can be easily proved, are all cancelled out
eventually. This result and the fact that any S-matrix element is
independent of the gauge parameter ( This is the so-called gauge-invariance
of S-matrix which is implied by the unitarity of S-matrix elements) indicate
that the task of renormalization for a gauge field theory is reduced to find
the running coupling constant and the running mass by their RGEs. These
running quantities completely describe the effect of higher order
perturbative corrections.

\setcounter{section}{3}

\section*{3. Ward Identity}

\setcounter{equation}{0}

In QED renormalization, the following Ward identity plays crucial role $^6$%
\begin{eqnarray}
\Lambda _\mu (p^{\prime };p)_{\mid p^{\prime }=p}=-\frac{\partial \Sigma (p)%
}{\partial p_\mu }
\end{eqnarray}
where $\Lambda _\mu (p^{\prime },p)$ represents the vertex correction which
is defined by taking out a coupling constant e and $\Sigma (p)$ denotes the
fermion self-energy. Firstly, we show how the above identity determines the
subtraction of the fermion self-energy $\sum (p)$. According to the Ward
identity, the divergence in $\Lambda _\mu (p^{\prime },p)$ should be, in the
GMS scheme, subtracted at a time-like (Minkowski) renormalization point for
the momenta of the external fermion lines, $p^2=p^{\prime 2}=\mu ^2$ which
implies $\not p=\not p^{\prime }=\mu $. When $\mu =m$ (the fermion mass), we
will come to the subtraction in the OS scheme. In the case of $\mu \neq m$,
the subtraction is defined on a generalized mass shell. At the
renormalization point $\mu $, we have 
\begin{eqnarray}
\Lambda _\mu (p^{\prime },p)\mid _{{\not p}^{\prime }={\not p}=\mu }=L\gamma
_\mu
\end{eqnarray}
Thus, the vertex correction may be represented as 
\begin{eqnarray}
\Lambda _\mu (p^{\prime },p)=L\gamma _\mu +\Lambda _\mu ^c(p^{\prime },p)
\end{eqnarray}
where $L$ is a divergent constant depending on $\mu $ and $\Lambda _\mu
^c(p^{\prime },p)$ is the finite correction satisfying the boundary
condition 
\begin{eqnarray}
\Lambda _\mu ^c(p^{\prime },p)\mid _{{\not p}^{\prime }={\not p}=\mu }=0
\end{eqnarray}

On inserting Eq.(3.3) into Eq.(3.1) and integrating the both sides of
Eq.(3.1) over the momentum $p^\mu $, we get 
\begin{eqnarray}
\Sigma (p)-\Sigma (\mu )=-({\not p}-\mu )L-\int_{p_0^\mu }^{p^\mu }dp^\mu
\Lambda _\mu ^c(p,p)
\end{eqnarray}
where the momentum $p_0$ is chosen to make ${\not p}_0=\mu $. Since the last
term on the RHS of Eq.(3.5) vanishes when $p^\mu \rightarrow p_0^\mu $, we
may write 
\begin{eqnarray}
\int_{p_0^\mu }^{p\mu }dp^\mu \Lambda _\mu ^c(p,p)=({\not p}-\mu )C(p^2)
\end{eqnarray}
where $C(p^2)$ is a convergent function satisfying the following boundary
condition 
\begin{eqnarray}
C(p^2)\mid _{p^2=\mu ^2}=0
\end{eqnarray}
which is implied by Eq.(3.4). Substituting Eq.(3.6) into Eq.(3.5) and
setting 
\begin{eqnarray}
\Sigma (\mu )=A
\end{eqnarray}
and 
\begin{eqnarray}
L=-B
\end{eqnarray}
the self-energy is finally written as 
\begin{eqnarray}
\Sigma (p)=A+({\not p}-\mu )[B-C(p^2)]
\end{eqnarray}
where the constants A and B have absorbed all the divergences appearing in
the $\Sigma (p)$. The above derivation shows that the subtraction given in
Eq.(3.10) is uniquely correct in the GMS scheme as it is compatible with the
Ward identity. According to the subtraction in Eq.(3.3), the full vertex can
be written as 
\begin{eqnarray}
\Gamma _\mu (p^{\prime },p) &=&\gamma _\mu +\Lambda _\mu (p^{\prime },p) 
\nonumber \\
&=&Z_1^{-1}\Gamma _\mu ^R(p^{\prime },p)
\end{eqnarray}
where $Z_1$ is the vertex renormalization constant defined as 
\begin{eqnarray}
Z_1^{-1}=1+L
\end{eqnarray}
and $\Gamma _\mu ^R(p^{\prime },p)$ is the renormalized vertex represented
by 
\begin{eqnarray}
\Gamma _\mu ^R(p^{\prime },p)=\gamma _\mu +\Lambda _\mu ^R(p^{\prime },p)
\end{eqnarray}
which satisfies the boundary condition 
\begin{eqnarray}
\Gamma \mu ^R(p^{\prime },p)\mid _{{\not p}^{\prime }={\not p}=\mu }=\gamma
_\mu
\end{eqnarray}

Based on the subtraction given in Eq.(3.10), the full fermion propagator may
be renormalized in such a way 
\begin{eqnarray}
iS_F(p)=\frac i{{\not p}-m-\Sigma (p)+i\varepsilon }=Z_2iS_F^R(p)
\end{eqnarray}
where $Z_2$ is the propagator renormalization constant defined by 
\begin{eqnarray}
Z_2^{-1}=1-B
\end{eqnarray}
and $S_F^R(p)$ denotes the renormalized propagator represented as 
\begin{eqnarray}
S_F^R(p)=\frac i{{\not p}-m_R-\Sigma _R(p)+i\varepsilon }
\end{eqnarray}
which has a boundary condition as follows 
\begin{equation}
S_F^R(p)\mid _{p^2=\mu ^2}=\frac i{{\not p}-m_R}
\end{equation}
In Eq.(3.17), $m_R$ and $\Sigma _R(p)$ designate the renormalized mass and
the finite correction of the self-energy respectively. The renormalized mass
is defined by 
\begin{eqnarray}
m_R=Z_m^{-1}m
\end{eqnarray}
where $Z_m$ is the mass renormalization constant expressed by 
\begin{eqnarray}
Z_m^{-1}=1+Z_2[Am^{-1}+(1-\mu m^{-1})B]
\end{eqnarray}
Particularly, from Eqs.(3.9), (3.12) and (3.16). it is clear to see 
\begin{eqnarray}
Z_1=Z_2
\end{eqnarray}
This just is the Ward identity obeyed by the renormalization constants.

~~Let us verify whether the Ward identity is fulfilled in the one-loop
approximation. The Feynman integrals of one-loop diagrams in QED have been
calculated in the literature by various regularization procedures$%
^{3-6,14,15},$. In the GMS scheme, the fermion self-energy depicted in
Fig.(1a), according to the dimensional regularization procedure, is
regularized in the form 
\begin{eqnarray}
\Sigma (p) &=&-\frac{e^2}{(4\pi )^2}(4\pi M^2)^\varepsilon \Gamma
(1+\varepsilon )\int_0^1dx\{\frac 1{\varepsilon \Delta (p)^\varepsilon
}[2(1-\varepsilon )  \nonumber \\
&&\times (1-x){\not p}-(4-2\varepsilon )m+(1-\xi )(m-2x{\not p})]-2(1-\xi ) 
\nonumber \\
&&\times (1-x)\frac{x^2p^2{\not p}}{\Delta (p)}\}
\end{eqnarray}
where $\varepsilon =2-\frac n2$, 
\begin{eqnarray}
\Delta (p)=p^2x(x-1)+m^2x
\end{eqnarray}
and M is an arbitrary mass introduced to make the coupling constant e to be
dimensionless in the space of dimension n. According to the definition shown
in Eq.(3.8) and noticing 
\begin{eqnarray}
p^2{\not p}=({\not p}-\mu )[p^2+\mu ({\not p}+\mu )]+\mu ^3
\end{eqnarray}
one can get from Eq.(3.22) 
\begin{eqnarray}
A &=&-\frac{e^2}{(4\pi )^2}(4\pi M^2)^\varepsilon \Gamma (1+\varepsilon
)\int_0^1dx\{\frac 1{\varepsilon \Delta (\mu )^\varepsilon }  \nonumber \\
&&\times [2\mu [1+(\xi -2)x-\varepsilon (1-x)]-(3+\xi -2\varepsilon )m] 
\nonumber \\
&&-2(1-\xi )(1-x)x^2\frac{\mu ^3}{\Delta (\mu )^{1+\varepsilon }}
\end{eqnarray}
where 
\begin{eqnarray}
\Delta (\mu )=x[\mu ^2(x-1)+m^2]
\end{eqnarray}
On substituting Eqs.(3.22) and (3.25) in Eq.(3.10), it is found that 
\begin{eqnarray}
B &=&[\Sigma (p)-A](p-\mu )^{-1}\mid _{{\not p}=\mu }  \nonumber \\
&=&-\frac{e^2}{(4\pi )^2}(4\pi M^2)^\varepsilon \Gamma (1+\varepsilon
)\int_0^1dx\{\frac 1{\varepsilon \Delta (\mu )^\varepsilon }[2(1-\varepsilon
)  \nonumber \\
&&\times (1-x)-2(1-\xi )x]+\frac{2\mu ^2}{\Delta (\mu )^{1+\varepsilon }}%
[2(1-\varepsilon )x(x-1)^2  \nonumber \\
&&+5(1-\xi )x^2(x-1)+\frac m\mu (3+\xi -2\varepsilon )x(x-1)]-\frac{4\mu ^4}{%
\Delta (\mu )^{2+\varepsilon }}  \nonumber \\
&&\times (1-\xi )(1+\varepsilon )(x-1)^2x^3\}
\end{eqnarray}

~~For the diagram of one-loop vertex correction shown in Fig.(1b), according
to the definition written in Eq.(3.2), it is not difficult to obtain, in the
n-dimensional space, the regularized form of the constant L 
\begin{eqnarray}
L &=&\frac{e^2}{(4\pi )^2}(4\pi M^2)^\varepsilon \Gamma (1+\varepsilon
)\int_0^1dx\{\frac{2x}{\varepsilon \Delta (\mu )^\varepsilon }[\varepsilon
(\varepsilon -\frac 32)  \nonumber \\
&&+\xi (1-\frac 12\varepsilon )]-\frac x{\Delta (\mu )^{1+\varepsilon
}}[2\mu ^2(\varepsilon -1)(x-1)^2  \nonumber \\
&&-(1-\xi )(x^2-x-1)\mu ^2+\varepsilon (1-\xi )(x-1)\mu ^2-4m\mu
[(2-\varepsilon )  \nonumber \\
&&\times (x-1)+\frac 12(1-\xi )(2-3x)+\frac 12\varepsilon (1-\xi
)(x-1)]+m^2[2(\varepsilon  \nonumber \\
&&-1)-(1-\varepsilon )(1-\xi )(x-1)]]+(1-\xi )\frac{(1+\varepsilon )}{\Delta
(\mu )^{2+\varepsilon }}(x-1)x^3\mu ^2  \nonumber \\
&&\times (m+\mu )^2\} \\
&&  \nonumber
\end{eqnarray}
With the expressions given in Eqs.(3.27) and (3.28), in the approximation of
order $e^2$, the renormalization constants defined in Eq.(3.12) and (3.16)
will be represented as $Z_1=1-L$ and $Z_2=1+B$. In the limit $\varepsilon
\to 0$, these constants are divergent, being not well-defined. So, to verify
the Ward identity in Eq.(3.21), it is suitable to see whether their
anomalous dimensions satisfy the corresponding identity 
\begin{equation}
\gamma _1=\gamma _2
\end{equation}
where ${\gamma _i(\mu )=\lim_{\varepsilon \rightarrow 0}\mu \frac d{d\mu
}\ln Z_i(\mu ,\varepsilon )}$ (i=1,2). For the renormalization group
calculations, in practice, the above identity is only necessary to be
required. Through direct calculation by using the constants in Eqs.(3.27)
and (3.28), it is easy to prove 
\begin{eqnarray}
\gamma _1 &=&\gamma _2  \nonumber \\
&=&-\frac{e^2}{(4\pi )^2}\{6\xi -6(3+\xi )\sigma +12\xi \sigma ^2+6(3+\xi
-2\xi \sigma )  \nonumber \\
&&\times \sigma ^3\ln \frac{\sigma ^2}{\sigma ^2-1}+4[2\xi -(3+\xi )\sigma
]\frac 1{\sigma ^2-1}
\end{eqnarray}
where ${\sigma =\frac m\mu }$. This identity guarantees the correctness of
the one-loop renormalizations in the GMS scheme. In the zero-mass limit $%
(\sigma \rightarrow 0)$, the identity in Eq.(3.30) reduces to the result
given in the MS scheme 
\begin{eqnarray}
\gamma _1=\gamma _2=\frac{\xi e^2}{8\pi ^2}
\end{eqnarray}
This result can directly be derived from such expressions of the constants B
and L that they are obtained from Eqs.(3.27) and (3.28) by setting m=0. It
is easy to see that in these expressions, only the terms proportional ${%
\varepsilon }^{-1}$ give nonvanishing contributions to the anomalous
dimensions. However, in the case of $m\neq 0$, the terms proportional to ${%
\varepsilon }^{-1}$ in Eqs.(3.27) and (3.28) give different results. In this
case, to ensure the identity in Eq.(3.29) to be satisfied, the other terms
without containing ${\varepsilon }^{-1}$ in Eqs.(3.27) and (3.28) must be
taken into account.

\setcounter{section}{4}

\section*{4.Effective Coupling Constant}

\setcounter{equation}{0}

~~ The RGE for the renormalized coupling constant may be immediately written
out from Eq.(2.2) by setting ${F=e,}$ 
\begin{eqnarray}
\mu \frac d{d\mu }e_R(\mu )+\gamma _e(\mu )e_R(\mu )=0
\end{eqnarray}
In the above, the anomalous dimension ${\gamma _e(\mu )}$ as defined in
Eq.(2.3) is now determined by the following renormalization constant$^5$

\begin{eqnarray}
Z_e=\frac{Z_1}{Z_2Z_3^{\frac 12}}=Z_3^{-\frac 12}
\end{eqnarray}
where the identity in Eq.(3.21) has been considered. The photon propagator
renormalization constant ${Z_3}$ is, in the GMS scheme, defined by 
\begin{eqnarray}
Z_3^{-1}=1+\Pi (\mu ^2)
\end{eqnarray}
where ${\Pi (\mu ^2})$ is the scalar function appearing in the photon
self-energy tensor ${\Pi _{\mu \nu }=(k}_\mu {k}_\nu {-k^2g_{\mu \nu })\Pi
(\mu ^2)}$. In view of Eq.(4.2), we can write 
\begin{eqnarray}
\gamma _e=\lim_{\varepsilon \to 0}\mu \frac d{d\mu }\ln Z_e=-\frac
12\lim_{\varepsilon \to 0}\mu \frac d{d\mu }\ln Z_3
\end{eqnarray}
For the one-loop diagram represented in Fig.(1c), according to Eq.(4.3), it
is easy to derive the regularized form of the constant ${Z_3}$ by the
dimensional regularization procedure 
\begin{eqnarray}
Z_3=1+\frac{e^2}{4\pi ^2}(4\pi M^2)^\varepsilon (2-\varepsilon )\frac{\Gamma
(1+\varepsilon )}\varepsilon \int_0^1\frac{dxx(x-1)}{[\mu
^2x(x-1)+m^2]^\varepsilon }
\end{eqnarray}
Substituting Eq.(4.5) into Eq.(4.4), it is found that 
\begin{eqnarray}
\gamma _e=-\frac{e^2}{12(\pi )^2}\{1+6\sigma ^2+\frac{12\sigma ^4}{\sqrt{%
1-4\sigma ^2}}\ln \frac{1+\sqrt{1-4\sigma ^2}}{1-\sqrt{1-4\sigma ^2}}\}
\end{eqnarray}
where $\sigma =\frac m\mu $. In this expression, the charge e and the mass m
are unrenormalized. In the approximation of order $e^2$, they can be
replaced by the renormalized ones $e_R$ and $m_R$ because in this
approximation, as pointed out in the previous literature$^{15}$, the lowest
order approximation of the relation between the $e(m)$ and the $e_R(m_R)$ is
only necessary to be taken into account. Furthermore, when we introduce the
scaling variable $\lambda $ for the renormalization point and set $\mu
_0=m_R $ (which can always be done since the ${\mu _0}$ is fixed, but may be
chosen at will), we have $\sigma =\frac{m_R}{\mu _0\lambda }=\frac 1\lambda $%
. Thus, with the expressions of Eq.(4.6), Eq.(4.1) may be rewritten in the
form 
\begin{eqnarray}
\lambda \frac{de_R(\lambda )}{d\lambda }=\beta (\lambda )
\end{eqnarray}
where 
\begin{eqnarray}
\beta (\lambda ) &=&-\gamma _e(\lambda )e_R(\lambda )  \nonumber \\
&=&\frac{e_R^3(\lambda )}{12\pi ^2}F_e(\lambda )
\end{eqnarray}
in which 
\begin{eqnarray}
F_e(\lambda )=1+\frac 6{\lambda ^2}+\frac{12}{\lambda ^4}f(\lambda )
\end{eqnarray}
\begin{eqnarray}
f(\lambda ) &=&\frac \lambda {\sqrt{\lambda ^2-4}}\ln \frac{\lambda +\sqrt{%
\lambda ^2-4}}{\lambda -\sqrt{\lambda ^2-4}}  \nonumber \\
&=&\cases{ \frac{2\lambda }{\sqrt{4-\lambda ^2}}cot^{-1}\frac \lambda
{\sqrt{4-\lambda ^2}},&if $ \lambda \leq 2$ \cr \frac{2\lambda
}{\sqrt{\lambda ^2-4}}coth^{-1}\frac \lambda {\sqrt{\lambda^2 -4}},&if $
\lambda \geq 2$\cr}
\end{eqnarray}
Upon substituting Eqs.(4.8)-(4.10) into Eq.(4.7) and then integrating
Eq.(4.7) by applying the familiar integration formulas, the effective
(running) coupling constant will be found to be 
\begin{eqnarray}
\alpha _R(\lambda )=\frac{\alpha _R}{1-\frac{2\alpha _R}{3\pi }G(\lambda )}
\end{eqnarray}
where ${\alpha _R(\lambda )=\frac{e_R^2(\lambda )}{4\pi }}$, $\alpha
_R=\alpha _R(1)$ and 
\begin{eqnarray}
G(\lambda ) &=&\int_1^\lambda \frac{d\lambda }\lambda F_e(\lambda ) 
\nonumber \\
&=&2+\sqrt{3}\pi -\frac 2{\lambda ^2}+(1+\frac 2{\lambda ^2})\frac 1\lambda
\varphi (\lambda )
\end{eqnarray}
in which 
\begin{eqnarray}
\varphi (\lambda ) &=&\sqrt{\lambda ^2-4}\ln \frac 12(\lambda +\sqrt{\lambda
^2-4})  \nonumber \\
&=&\cases{ -\sqrt{4-\lambda^2 }\cos ^{-1}\frac \lambda 2,&if $ \lambda \leq
2$ \cr \sqrt{\lambda ^2-4}\cosh ^{-1}\frac \lambda 2,&if $ \lambda \geq
2$\cr}
\end{eqnarray}
As mentioned in Sect.2, the variable $\lambda $ is also the scaling
parameter of momenta, $p=\lambda p_0$ and it is convenient to put ${p_0}^2={%
\mu _0}^2$ so as to apply the boundary condition. Thus, owing to the choice $%
\mu _0=m_R$, we have ${p_0}^2={m_R}^2$ and $\lambda =(\frac{p^2}{m_R^2}%
)^{\frac 12}$. In this case, it is apparent that when $\lambda =1$,
Eq.(4.11) will be reduced to the result given on the mass shell, $\alpha
_R(1)=\alpha _R=\frac 1{137}$ which is identified with that as measured in
experiment.

~~ The behavior of the $\alpha _R(\lambda )$ are exhibited in Figs.(2) and
(3). For small $\lambda $, Eq.(4.11) may be approximated by 
\begin{eqnarray}
\alpha _R(\lambda )\approx \frac 34\lambda ^3
\end{eqnarray}
It is clear that when $\lambda \rightarrow 0$, the $\alpha _R(\lambda )$
tends to zero. This desirable behavior, which indicates that at large
distance (small momentum) the interacting particles decouple, is completely
consistent with our knowledge about the electromagnetic interaction. For
large momentum (small distance), Eq.(4.11) will be approximated by 
\begin{eqnarray}
\alpha _R(\lambda )\approx \frac{\alpha _R}{1-\frac{2\alpha _R}{3\pi }\ln
\lambda }
\end{eqnarray}
This result was given previously in the mass-independent MS scheme. In the
latter scheme, the $\beta -$ function in Eq.(4.8) is only a function of the $%
e_R(\lambda )$ since $F_e(\lambda )=1$ due to $m=0$ in this case. But, in
general, the mass of a charged particle is not zero. Therefore, the result
given in the MS scheme can only be viewed as an approximation in the large
momentum limit from the viewpoint of conventional perturbation theory.
Fig.(3) shows that the $\alpha _R(\lambda )$ increases with the growth of $%
\lambda $ and tends to infinity when the $\lambda $ approaches an extremely
large value $\lambda _0\approx e^{\frac{2\pi }{3\alpha _R}}=e^{287}$ (the
Landau pole). If the $\lambda $ goes from $\lambda _0$ to infinity, we find,
the $\alpha _R(\lambda )$ will become negative and tends to zero. This
result is unreasonable, conflicting with the physics. The unreasonableness
indicates that in the region $[\lambda _0,\infty )$, the QED perturbation
theory and even the QED itself is invalid$^5$.

\setcounter{section}{5}

\section*{5.Effective Fermion Mass}

\setcounter{equation}{0}

~~The RGE for a renormalized fermion mass can be directly read from Eq.(2.2)
when we set $F=m_R$. Noticing $\mu \frac d{d\mu }=\lambda \frac d{d\lambda }$%
, this equation may be written in the form 
\begin{eqnarray}
\lambda \frac{dm_R(\lambda )}{d\lambda }=-\gamma _m(\lambda )m_R(\lambda )
\end{eqnarray}
where the anomalous dimension $\gamma _m(\lambda )$, according to the
definition in Eq.(2.3), can be derived from the renormalization constant
represented in Eq.(3.20). At one-loop level, by making use of the constants
A, B and $Z_2$ which were written in Eqs.(3.25), (3.27) and (3.16)
respectively, in the approximation of order $e^2$, it is not difficult to
derive 
\begin{eqnarray}
\gamma _m(\lambda )=\lim_{\varepsilon \to 0}\mu \frac d{d\mu }\ln Z_m=\frac{%
e_R^2}{(4\pi )^2}F_m(\lambda )
\end{eqnarray}
where 
\begin{eqnarray}
F_m(\lambda ) &=&2\xi \lambda +6[3+2\xi -\frac{3(1+\xi )}\lambda +\frac{2\xi 
}{\lambda ^2}]  \nonumber \\
&&\ -\frac{12(1+\xi )\lambda }{1+\lambda }+6[3+\xi -\frac{3(1+\xi )}\lambda +%
\frac{2\xi }{\lambda ^2}]\frac 1{\lambda ^2}\ln \left| 1-\lambda ^2\right|
\end{eqnarray}
here the relation $\sigma =\frac 1\lambda $ has been used. Inserting
Eq.(5.2) into Eq.(5.1) and integrating the latter equation, one may obtain 
\begin{eqnarray}
m_R(\lambda )=m_Re^{-S(\lambda )}
\end{eqnarray}
This just is the effective (running) fermion mass where $m_R=m_R(1)$ which
is given on the mass-shell and 
\begin{eqnarray}
S(\lambda )=\frac 1{4\pi }\int_1^\lambda \frac{d\lambda }\lambda \alpha
_R(\lambda )F_m(\lambda )
\end{eqnarray}
In the above, the bare charge appearing in Eq.(5.2) has been replaced by the
renormalized one and further by the running one shown in Eq.(4.11). If the
coupling constant in Eq.(5.5) is taken to be the constant defined on the
mass-shell, the integral over $\lambda $ can be explicitly calculated. The
result is 
\begin{eqnarray}
S(\lambda )=\frac{\alpha _R}{4\pi }[\varphi _1(\lambda )+\xi \varphi
_2(\lambda )]
\end{eqnarray}
where 
\begin{eqnarray}
\varphi _1(\lambda )=3(1-\lambda )\{\frac 2\lambda +[\frac 2{\lambda
^3}-\frac 1{\lambda ^2}(1+\lambda )]\ln \left| 1-\lambda ^2\right| \}
\end{eqnarray}
and 
\begin{eqnarray}
\varphi _2(\lambda ) &=&2\lambda +5\ln \lambda -\frac{20}3\ln \frac
12(1+\lambda )-\frac{38}{3\lambda }  \nonumber \\
&&\ -\frac{11}{2\lambda ^2}-\frac{55}6+[\frac 56-\frac{17}{6\lambda ^2}%
+\frac 5{2\lambda ^3}-\frac 1{2\lambda ^4}]  \nonumber \\
&&\ \times \ln \left| 1-\lambda ^2\right|
\end{eqnarray}
in which 
\begin{eqnarray}
\ln \left| 1-\lambda ^2\right| =\cases{ 2[\ln (1+\lambda )-\tanh
^{-1}\lambda ],&if $\lambda \leq 1 $ \cr 2[\ln (1+\lambda )-coth^{-1}\lambda
],&if $\lambda \geq 1 $\cr}
\end{eqnarray}
As we see, the function $S(\lambda )$ and hence the effective mass $%
m_R(\lambda )$ are gauge-dependent. The gauge-dependence is displayed in
Fig.(4). The figure shows that for $\xi <10,$ the effective masses given in
different gauges are almost the same and behave as a constant in the region
of $\lambda <1$, while, in the region of $\lambda >1,$ they all tend to zero
with the growth of $\lambda $. But, the $m_R(\lambda )$ given in the gauge
of $\xi \neq 0$ goes to zero more rapidly than the one given in the Landau
gauge $(\xi =0)$. To be specific, in the following we show the result given
in the Landau gauge which was regarded as the preferred gauge in the
literature$^{3,29}$, 
\begin{eqnarray}
m_R(\lambda )=m_Re^{-\frac{\alpha _R}{4\pi }\varphi _1(\lambda )}
\end{eqnarray}
In the limit $\lambda \rightarrow 0$, 
\begin{eqnarray}
m_R(\lambda )\rightarrow m_Re^{-\frac{3\alpha _R}{4\pi }}=1.001744m_R
\end{eqnarray}
This clearly indicates that when $\lambda $ varies from 1 to zero, the $%
m_R(\lambda )$ almost keeps unchanged. This result physically is reasonable.
Whereas, in the region of $\lambda >1$, the $m_R(\lambda )$ decreases with
increase of $\lambda $ and goes to zero near the critical point $\lambda _0$%
. This behavior suggests that at very high energy, the fermion mass may be
ignored in the evaluation of S-matrix elements.

\setcounter{section}{6}

\section*{6.Comments and Discussions}

\setcounter{equation}{0}

~~In this paper, the QED renormalization has been restudied in the GMS
scheme. The exact and explicit expressions of the one-loop effective
coupling constant and fermion mass are obtained in the mass-dependent
renormalization scheme and show reasonable asymptotic behaviors. A key point
to achieve these results is that the subtraction is performed in the way of
respecting the Ward identity, i.e. the gauge symmetry. For comparison, it is
mentioned that in some previous literature$^{15,19}$, the fermion
self-energy is represented in such a form 
\begin{eqnarray}
\Sigma (p)=A(p^2)\not p+B(p^2)m
\end{eqnarray}
If we subtract the divergence in the $\Sigma (p)$ at the renormalization
point $p^2=\mu ^2$, the fermion propagator is still expressed in the form as
written in Eqs.(3.15) and (3.17); but, the renormalization constants $Z_2$
and $Z_m$ are now defined by 
\begin{eqnarray}
Z_2^{-1}=1-A(\mu ^2),Z_m^{-1}=Z_2[1+B(\mu ^2)]
\end{eqnarray}
The one-loop expressions of the $A(\mu ^2)$ and $B(\mu ^2)$ can directly be
read from Eq.(3.22). Here, we show the one-loop anomalous dimension of
fermion propagator which is given by the above subtraction 
\begin{eqnarray}
\gamma _2 &=&\lim_{\varepsilon \to 0}\mu \frac d{d\mu }\ln Z_2  \nonumber \\
&=&\frac{\xi e^2}{4\pi ^2}[\frac 12+\sigma ^2-\sigma ^4\ln \frac{\sigma ^2}{%
\sigma ^2-1}]
\end{eqnarray}
In comparison of the above $\gamma _2$ with the $\gamma _1$ shown in
Eq.(3.30), we see, the Ward identity in Eq.(3.29) can not be fulfilled
unless in the zero-mass limit $(\sigma \to 0)$. Since the Ward identity is
an essential criterion to identify whether a subtraction is correct or not,
the subtraction stated above should be excluded from the mass-dependent
renormalization.

~~Another point we would like to address is that in the Ward identity shown
in Eq.(3.1), the momenta p and $p^{\prime }$ on the fermion lines in the
vertex are set to be equal. According to the energy-momentum conservation,
the momentum k on the photon line should be equal to zero. correspondingly,
the subtraction shown in Eq.(3.2) was carried out at the so-called
asymmetric points, $p^{\prime }{}^2=p^2=\mu ^2$ and $k^2=(p^{\prime }-p)^2=0$%
. These subtraction points coincide with the energy- momentum conservation,
i.e., the Lorentz-invariance. Nevertheless, the subtraction performed at the
symmetric point $p^{\prime 2}=p^2=k^2=-\mu ^2$ was often used in the
previous works $^{3,19,20}$. This subtraction not only makes the calculation
too complicated, but also violates the energy- momentum conservation which
holds in the vertex. That is why the symmetric point subtraction is beyond
our choice.

~~As mentioned in Introduction, the GMS subtraction is a kind of MOM scheme
in which the renormalization points are chosen to be time-like. In contrast,
in the conventional MOM scheme$^{3,6,29}$, the renormalization points were
chosen to be space-like, i.e. $p_i^2=-\mu ^2$ which implies $\not p_i=i\mu $%
. In this scheme, the one-loop result for the anomalous dimension $\gamma _e$
can be written out from Eq.(4.6) by the transformation $\sigma \to -i\sigma $%
. That is 
\begin{eqnarray}
\gamma _e=-\frac{4e^2}{3(4\pi )^2}\{1-6\sigma ^2+\frac{12\sigma ^4}{\sqrt{%
1+4\sigma ^2}}\ln \frac{\sqrt{1+4\sigma ^2}+1}{\sqrt{1+4\sigma ^2}-1}\}
\end{eqnarray}
which is identical to that given in Refs.(6) and (29). In comparison of
Eq.(6.4) with Eq.(4.6), we see, the coefficients of $\sigma ^2$ in Eq.(6.4)
changes a minus sign. Substituting Eq.(6.4) into Eq.(4.1) and solving the
latter equation, we obtain the running coupling constant which is still
represented in Eq.(4.11), but the function $G(\lambda )$ in Eq.(4.11) is now
given by 
\begin{eqnarray}
G(\lambda ) &=&\frac 2{\lambda ^2}-2-\frac{\sqrt{\lambda ^2+4}}\lambda
(\frac 2{\lambda ^2}-1)\ln \frac 12(\lambda +\sqrt{\lambda ^2+4})  \nonumber
\\
&&+\sqrt{5}\ln \frac 12(1+\sqrt{5})
\end{eqnarray}
This expression is obviously different from the corresponding one written in
Eqs.(4.12) and (4.13). At small distance $\left( \lambda \rightarrow \infty
\right) $, Eq.(6.5) still gives the approximate expression presented in
Eq.(4.15). However, at large distance, as shown in Fig.(2), the $\alpha
_R(\lambda )$ behaves almost a constant. When $\lambda \rightarrow 0$, it
approaches to a value equal to 0.99986$\alpha _R$, unlike the $\alpha
_R(\lambda )$ given in Eqs.(4.11)-(4.13) which tends to zero. Let us examine
the effective mass. As we have seen from Sect.5, the one-loop effective
fermion mass given in the GMS scheme is real. However, in the space-like
momentum subtraction, due to $\not p=i\mu $, the effective mass will contain
an imaginary part. This result can be seen from the function $F_m(\lambda )$
whose one-loop expression given in the usual MOM scheme can be obtained from
Eq.(5.3) by the transformation $\lambda \to i\lambda $ and therefore becomes
complex. The both of subtractions may presumably be suitable for different
processes of different physical natures. But, if the effective mass is
required to be real, the subtraction at space-like renormalization point
should also be ruled out.

~~As pointed out in Sect.4, the effective coupling constant shown in
Eq.(4.15) which was obtained in the MS scheme is only an approximation given
in the large momentum limit from the viewpoint of conventional perturbation
theory. Why say so? As is well-known, the MS scheme is a mass-independent
renormalization scheme in which the fermion mass is set to vanish in the
process of subtraction. The reasonability of this scheme was argued as
follows$^{1,15}$ .The fermion propagator can be expanded as a series 
\begin{equation}
\frac 1{\not p-m}=\frac 1{\not p}+\frac 1{\not p}m\frac 1{\not p}+\frac 1{%
\not p}m\frac 1{\not p}m\frac 1{\not p}+\cdot \cdot \cdot
\end{equation}
According to this expansion, the massive propagator $\frac 1{\not p-m}$ may
be replaced by the massless one $\frac 1{\not p}.$ At the same time, the
fermion mass, as the coupling constant, can also be treated as an expansion
parameter for the perturbation series. Nevertheless, in the mass-dependent
renormalization as shown in this paper, the massive fermion propagator is
employed in the calculation and only the coupling constant is taken to be
the expansion parameter of the perturbation series. Thus, in order to get
the perturbative result of a given order of the coupling constant in the
mass-dependent renormalization, according to Eq.(6.6), one has to compute an
infinite number of terms in the MS scheme. If only the first term in
Eq.(6.6) is considered in the MS scheme, the result derived in the this
scheme, comparing to the corresponding one obtained in the mass-dependent
renormalization, can only be viewed as an approximation given in the large
momentum limit . Even if in this limit, a good renormalization scheme should
still be required to eliminate the ambiguity and give an unique result. To
this end, we may ask whether there should exist the difference between the
MS scheme and the $\overline{MS}$ scheme ${^2}$? As one knows, when the
dimensional regularization is employed in the mass-independent
renormalization, the MS scheme only subtracts the divergent term having the $%
\varepsilon $-pole in a Feynman integral which is given in the limit $%
\varepsilon \rightarrow 0$ and uses this term to define the renormalization
constant. While, the $\overline{MS}$ scheme is designed to include the
unphysical terms ${\gamma -\ln 4\pi }$ (here $\gamma $ is the Euler
constant) in the definition of the renormalization constant. The unphysical
terms arise from a special analytical continuation of the space-time
dimension from n to 4. When the two different renormalization constants
mentioned above are inserted into the relation $e=Z_3^{-\frac 12}e_R$, one
would derive a relation between the two different renormalized coupling
constants given in the MS and $\overline{MS}$ schemes if the higher order
terms containing the $\varepsilon $-pole are ignored. It would be pointed
out that the above procedure of leading to the difference between the MS and 
$\overline{MS}$ schemes is not appropriate because the procedure is based on
direct usage of the divergent form of the renormalization constants. As
emphasized in the Introduction, according to the convergence principle, it
is permissible to use such renormalization constants to do a meaningful
calculation. The correct procedure of deriving a renormalized quantity is to
solve its RGE whose solution is uniquely determined by the anomalous
dimension (other than the renormalization constant itself) and boundary
condition. In computing the anomalous dimension, the rigorous procedure is
to start from the regularized form of the renormalization constant. In the
regularized form, it is unnecessary and even impossible to divide a
renormalization constant into a divergent part and a convergent part. Since
the anomalous dimension is a convergent function of $\varepsilon $ due to
that the factor $\frac 1\varepsilon $ disappears in it, the limit $%
\varepsilon \rightarrow 0$ taken after the differentiation with respect to
the renormalization point would give an unambiguous result. Especially, the
unphysical factor $\left( 4\pi \right) ^\varepsilon \Gamma (1+\varepsilon )$
appearing in Eqs.(3.25), (3.27) and (4.5) straightforwardly approaches 1 in
the limit. Therefore, the unphysical terms $\gamma -\ln 4\pi $ could not
enter the anomalous dimension and the effective coupling constant. Even if
we work in the zero-mass limit or in the large momentum regime, we have an
only way to obtain the effective coupling constant as shown in Eq.(4.15) in
the one-loop approximation. That is to say, it is impossible, in this case,
to result in the difference between the MS and $\overline{MS}$ schemes and
also the difference between the MOM and the MS schemes.

The above discussions suggest that the ambiguity arising from different
renormalization prescriptions may be eliminated by the necessary physical
and mathematical requirements as well as the boundary conditions. It is
expected that the illustration given in this paper for the QED one-loop
renormalization performed in a mass-dependent scheme would provide a clue on
how to do the QED multi-loop renormalization and how to give an improved
result for the QCD renormalization.

\begin{center}
{\bf Acknowledgment}
\end{center}

The authors wish to thank professor Shi-Shu Wu for useful discussions. This
project was supposed in part by National Natural Science Foundation.

\renewcommand{\thesection}{\Alph{section}}\setcounter{section}{1}%
\setcounter{section}{1}\setcounter{equation}{0}

\begin{center}
{\bf Appendix: Illustration of the regularization procedure by a couple of
mathematical examples}
\end{center}

We believe that if a quantum field theory is built up on the faithful basis
of physical principles and really describes the physics, a S-matrix element
computed from such a theory is definite to be convergent even though there
occur divergences in the perturbation series of the matrix element. The
occurrence of divergences in the perturbation series, in general, is not to
be a serious problem in mathematics. But, to compute such a series, it is
necessary to employ an appropriate regularization procedure. For example,
for the following convergent integral 
\begin{equation}
f(a)=\int_0^\infty dxe^{-ax}
\end{equation}
which equals to $\frac 1a$, if we evaluate it by utilizing the series
expansion of the exponential function 
\begin{equation}
e^{-ax}=\sum_{n=0}^\infty \frac{(-a)^n}{n!}x^n
\end{equation}
as we see, the integral of every term in the series is divergent. In this
case, interchange of the integration with the summation actually is not
permissible. When every integral in the series is regularized, the
interchange is permitted and all integrals in the series become calculable.
Thus, the correct procedure of evaluating the integral by using the series
expansion is as follows 
\[
f(a)=\lim_{\Lambda \rightarrow \infty }\sum_{n=0}^\infty \frac{(-a)^n}{n!}%
\int_0^\Lambda dxx^n 
\]
\begin{eqnarray}
\ &=&\lim_{\Lambda \rightarrow \infty }\sum_{n=0}^\infty \frac{(-a)^n}{n!}%
\frac{\Lambda ^{n+1}}{n+1}=\lim_{\Lambda \rightarrow \infty }\frac
1a(1-e^{-a\Lambda })  \nonumber \\
\ &=&\frac 1a
\end{eqnarray}
This example is somewhat analogous to the perturbation series in the quantum
field theory and suggests how to do the calculation of the series with the
help of a regularization procedure. Unfortunately, in practice, we are not
able to compute all the terms in the perturbation series. In this situation,
we can only expect to get desired physical results from a finite order
perturbative calculation. How to do it? To show the procedure of such a
calculation, let us look at another mathematical example. The following
integral 
\begin{equation}
F(a)=\int_0^\infty dx\frac{e^{-(x+a)}}{(x+a)^2}
\end{equation}
where $a>0$ is obviously convergent. When the exponential function is
expanded as the Taylor series, the integral will be expressed as 
\begin{eqnarray}
F(a) &=&\int_0^\infty dx\sum_{n=0}^\infty \frac{(-1)^n}{n!}%
(x+a)^{n-2}=\int_0^\infty \frac{dx}{(x+a)^2}-\int_0^\infty \frac{dx}{x+a} 
\nonumber \\
&&\ \ +\int_0^\infty dx\sum_{n=2}^\infty \frac{(-1)^n}{n!}(x+a)^{n-2}
\end{eqnarray}
Clearly, in the above expansion, the first term is convergent, similar to
the tree-approximate term in the perturbation theory of a quantum field
theory, the second term is logarithmically divergent, analogous to the
one-loop-approximate term and the other terms amount to the higher order
corrections in the perturbation theory which are all divergent. To calculate
the integral in Eq.(A.4), it is convenient at first to evaluate its
derivative, 
\begin{equation}
\frac{dF(a)}{da}=-\int_0^\infty dx\frac{(x+a+2)}{(x+a)^3}e^{-(x+a)}
\end{equation}
which is equal to $-e^{-a}/a^2$ as is easily seen from integrating it over x
by part. In order to get this result from the series expansion, we have to
employ a regularization procedure. Let us define 
\begin{equation}
F_\Lambda (a)=\int_0^\Lambda \frac{e^{-(x+a)}}{(x+a)^2}
\end{equation}
Then, corresponding to Eq.(A.5), we can write 
\begin{eqnarray}
\frac{dF_\Lambda (a)}{da} &=&-2\int_0^\Lambda \frac{dx}{(x+a)^3}%
+\int_0^\Lambda \frac{dx}{(x+a)^2}+\sum_{n=2}^\infty \frac{(-1)^n}{n!}%
(n-2)\int_0^\Lambda dx(x+a)^{n-3}  \nonumber \\
\ &=&\sum_{n=0}^\infty \frac{(-1)^n}{n!}[(\Lambda +a)^{n-2}-a^{n-2}]=\frac{%
e^{-(\Lambda +a)}}{(\Lambda +a)^2}-\frac{e^{-a}}{a^2}
\end{eqnarray}
From the above result, it follows that 
\begin{equation}
\frac{dF(a)}{da}=\lim_{\Lambda \rightarrow \infty }\frac{dF_\Lambda (a)}{da}%
=-\frac{e^{-a}}{a^2}
\end{equation}
This is the differential equation satisfied by the function F(a). Its
solution can be expressed as 
\begin{equation}
F(a)=F(a_0)-\int_{a_0}^ada\frac{e^{-a}}{a^2}
\end{equation}
where $a_0$ $>0$ is a fixed number which should be determined by the
boundary condition of the equation (A.9). Now, let us focus our attention on
the second integral in the first equality of Eq.(A.8). This integral is
convergent in the limit $\Lambda \rightarrow \infty $ and can be written as 
\begin{equation}
\frac{dF_1(a)}{da}=\lim_{\Lambda \rightarrow \infty }\int_0^\Lambda dx\frac
1{(x+a)^2}=\int_0^\infty dx\frac 1{(x+a)^2}=\frac 1a
\end{equation}
Integrating the above equation over a, we get 
\begin{equation}
F_1(a)=F_1(a_0)+\ln \frac a{a_0}
\end{equation}
If setting 
\begin{equation}
F_1(a_0)=\ln \frac{a_0}\mu
\end{equation}
where $\mu $ is a finite number, Eq.(A.12) becomes 
\begin{equation}
F_1(a)=\ln \frac a\mu
\end{equation}
This result is finite as long as the parameter $\mu $ is not taken to be
zero and can be regarded as the contribution of the divergent integral $%
F_1(a)$ appearing in the second term of Eq.(A.5) to the convergent integral
F(a). The procedure described above for evaluating the function F(a) much
resembles the renormalization group method and the approach proposed in
Ref.(9). It shows us how to calculate a finite quantity from its series
expansion which contains divergent integrals.

\section{FIGURE CAPTIONS}

Fig.(1) The one-loop diagrams.

Fig.(2) The one-loop effective coupling constants given in the region of
small momenta. The solid curve represents the result obtained at time-like
subtraction point. The dashed curve represents the one given at space-like
subtraction point.

Fig.(3) The one-loop effective coupling constants for large momenta. The
both curves represent the same ones as in Fig.(2).

Fig.(4) The one-loop effective electron masses given in the Landau gauge and
some other gauges.

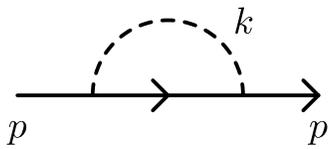
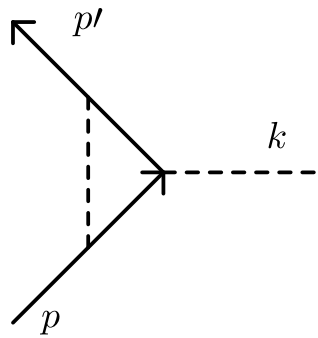
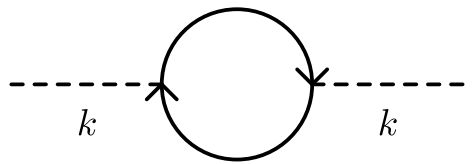
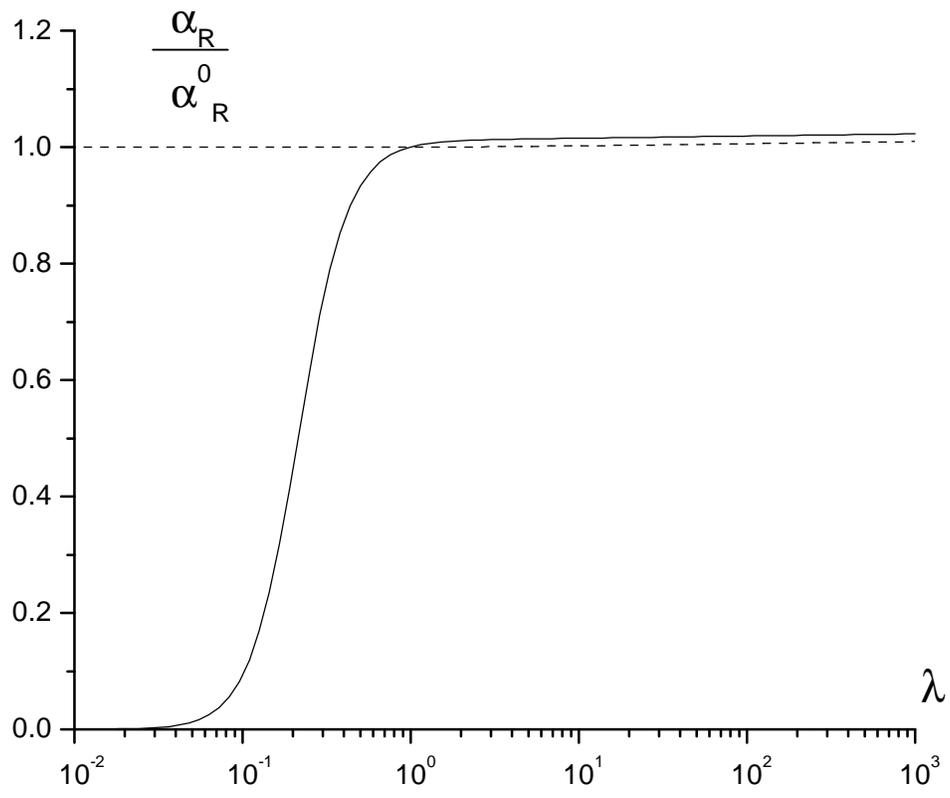
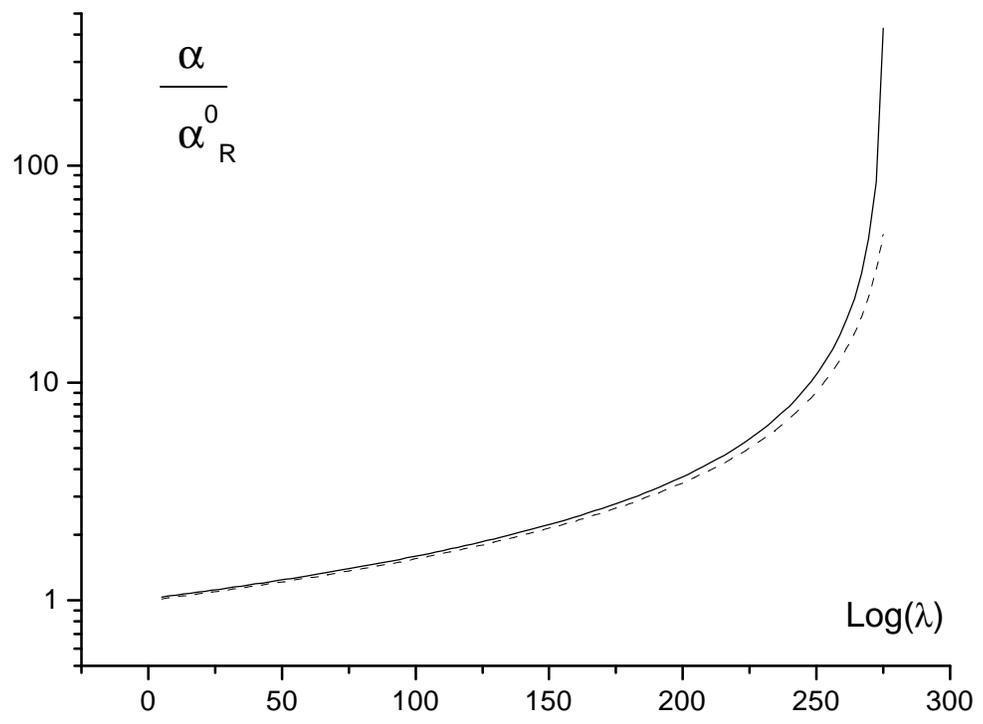
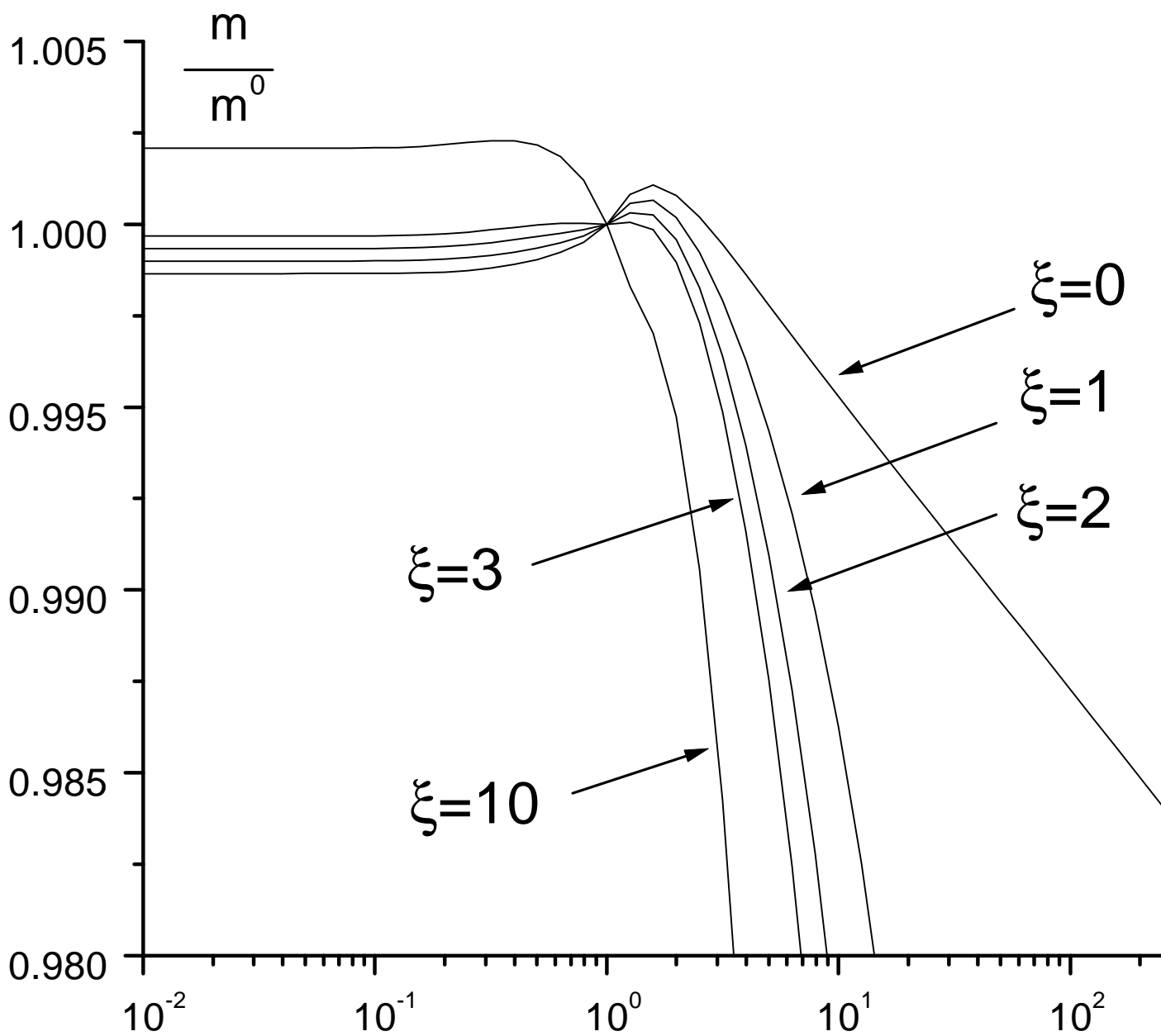
\end{document}